\documentclass[
reprint,
superscriptaddress,
amsmath,
amssymb,
prx,
]{revtex4-2}

\usepackage{graphicx}  
\usepackage{hyperref}  
\usepackage{siunitx}
\usepackage{xspace}

\newcommand{\ket}[1]{\ensuremath{\lvert #1 \rangle}\xspace}

\newcommand{\avg}[1]{\ensuremath{\langle #1 \rangle}\xspace}
\newcommand{\figref}[2][]{Fig.\,\ref{#2}#1}
\newcommand{\eqnref}[1]{Eq.\,(\ref{#1})}

\makeatletter
\def\maketitle{
\@author@finish
\title@column\titleblock@produce
\suppressfloats[t]}
\makeatother

\begin{document}

\newcommand{\papertitle}{Observation of brane parity order in programmable optical lattices}

\title{\papertitle{}}

\author{David Wei}
\affiliation{Max-Planck-Institut f\"{u}r Quantenoptik, 85748 Garching, Germany}
\affiliation{Munich Center for Quantum Science and Technology (MCQST), 80799 Munich, Germany}
\author{Daniel Adler}
\affiliation{Max-Planck-Institut f\"{u}r Quantenoptik, 85748 Garching, Germany}
\affiliation{Munich Center for Quantum Science and Technology (MCQST), 80799 Munich, Germany}
\author{Kritsana Srakaew}
\affiliation{Max-Planck-Institut f\"{u}r Quantenoptik, 85748 Garching, Germany}
\affiliation{Munich Center for Quantum Science and Technology (MCQST), 80799 Munich, Germany}
\author{Suchita Agrawal}
\affiliation{Max-Planck-Institut f\"{u}r Quantenoptik, 85748 Garching, Germany}
\affiliation{Munich Center for Quantum Science and Technology (MCQST), 80799 Munich, Germany}
\author{Pascal Weckesser}
\affiliation{Max-Planck-Institut f\"{u}r Quantenoptik, 85748 Garching, Germany}
\affiliation{Munich Center for Quantum Science and Technology (MCQST), 80799 Munich, Germany}
\author{Immanuel Bloch}
\affiliation{Max-Planck-Institut f\"{u}r Quantenoptik, 85748 Garching, Germany}
\affiliation{Munich Center for Quantum Science and Technology (MCQST), 80799 Munich, Germany}
\affiliation{Fakult\"{a}t f\"{u}r Physik, Ludwig-Maximilians-Universit\"{a}t, 80799 Munich, Germany}
\author{Johannes Zeiher}
\affiliation{Max-Planck-Institut f\"{u}r Quantenoptik, 85748 Garching, Germany}
\affiliation{Munich Center for Quantum Science and Technology (MCQST), 80799 Munich, Germany}

\date{\today}

\begin{abstract}
The Mott-insulating phase of the two-dimensional (2d) Bose-Hubbard model is expected to be characterized by a non-local brane parity order.
Parity order captures the presence of microscopic particle-hole fluctuations and entanglement, whose properties depend on the underlying lattice geometry.
We realize 2d Bose-Hubbard models in dynamically tunable lattice geometries, using neutral atoms in a novel passively phase-stable tunable optical lattice in combination with programmable site-blocking potentials.
We benchmark the performance of our system by single-particle quantum walks in the square, triangular, kagome and Lieb lattice.
In the strongly correlated regime, we microscopically characterize the geometry dependence of the quantum fluctuations and experimentally validate the brane parity as a proxy for the non-local order parameter signaling the superfluid-to-Mott insulating phase transition.
\end{abstract}

\maketitle

\section{Introduction}
According to seminal work by Landau, second-order phase transitions are signaled by a change of a local order parameter.
However, some phase transitions defy this classification in terms of a local order parameter, and require, as a generalization, non-local order parameters to describe their underlying structure~\cite{dennijs1989,kennedy1992}.
The Haldane insulator constitutes a celebrated example for such a phase, in which a string correlator captures the underlying non-local hidden order~\cite{haldane1983,dallatorre2006}, which has recently also been realized experimentally~\cite{deleseleuc2019,sompet2022}.
Interestingly, the Mott-insulating (MI) phase of the Bose-Hubbard (BH) model also features non-local order, which accounts for quantum fluctuations in the form of bound particle-hole pairs~\cite{endres2011,rath2013,hartke2020}.
In one-dimensional (1d) BH chains, the MI order has been revealed by a parity order parameter of the on-site occupation~\cite{berg2008,endres2011,rath2013}.
In two dimensions (2d) the brane parity was proposed as a generalization of parity order for square lattices~\cite{degliespostiboschi2016,fazzini2017}.
However, up to now, experiments directly measuring the brane parity in any 2d lattice geometry are lacking, as well as its experimental validation as an order parameter for the MI phase in 2d.
A strategy to achieve the latter is provided by mean-field theory, which predicts that the location of the phase transition should scale with the coordination number and thus the underlying lattice geometry.
This scaling was explicitly probed by measuring the local order parameter in the SF phase~\cite{thomas2017}.
Observing such scaling also in the brane parity provides an indication for the suitability of the brane parity as 2d non-local order parameter.

Neutral atoms in optical lattices provide a pristine testbed to realize low-dimensional Hubbard models~\cite{gross2017} and offer techniques for the detection of local observables using quantum gas microscopes~\cite{bakr2009,sherson2010}.
Optical lattices arise through the interference pattern of laser beams, whose layout is carefully chosen for a specific target geometry and has led to the realization of a variety of lattices~\cite{tarruell2012,jo2012,taie2015,yamamoto2020,yang2021}.
While optical lattices benefit from their inherent homogeneity and stability, the static nature of a given beam layout restricts systems to fixed spatial geometries and makes dynamical changes within a single experimental run challenging.
In contrast, arrays of optical tweezers can be generated in almost freely programmable geometries~\cite{barredo2016} and have allowed for studies of a variety of many-body spin models.
A number of approaches have been brought forward to allow for similar programmability for itinerant atoms based on realizing small systems of tunnel-coupled optical tweezers~\cite{kaufman2014,spar2022} or dynamically controllable lattices~\cite{tarruell2012}.
However, tweezer arrays in the itinerant regime are difficult to scale to large system sizes due to inhomogeneities and concomitantly a large calibration overhead, and the dynamical control over lattices typically involves complex active phase stabilization techniques~\cite{tarruell2012,xu2022}.

Here, we report on the realization of 2d Bose-Hubbard models in passively phase-stable optical lattices with square or triangular base geometry, which we combine with local site-blocking beams to realize programmable unit cells.
We demonstrate this novel degree of flexible control by implementing square, triangular, kagome and Lieb lattices in one experimental setup, and benchmark their quality through single-particle quantum walks.
Increasing the atomic density, we microscopically probe the strongly interacting regime through non-local quantum fluctuations and show their dependence on the underlying lattice structure.
Our measurements provide a quantitative characterization of the phase transition point in these lattices and experimentally establish the brane parity~\cite{rath2013,degliespostiboschi2016,fazzini2017} as a meaningful non-local observable to characterize the SF-MI phase transition in 2d models.

\begin{figure*}
\centering
\includegraphics{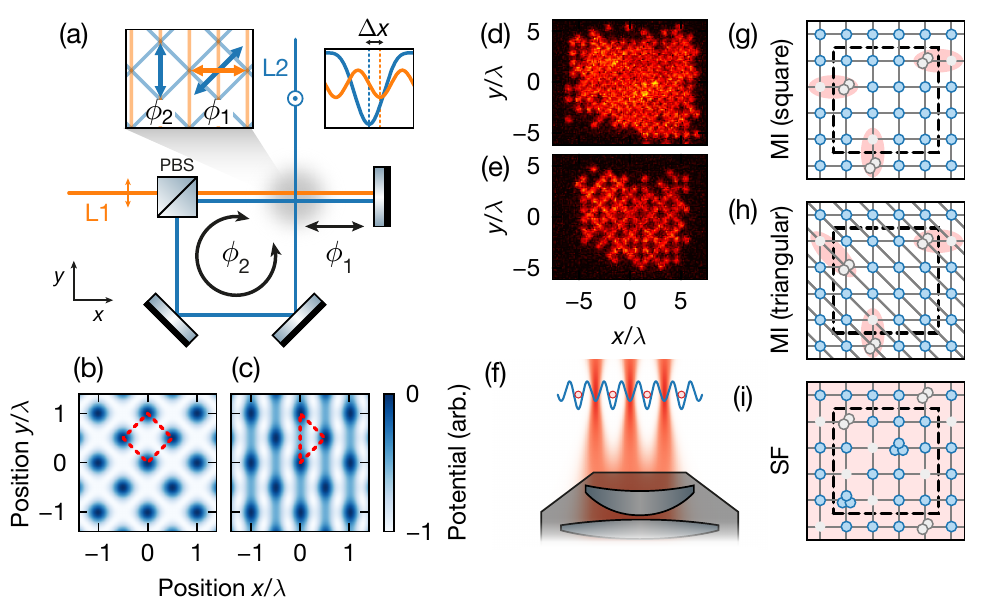}
\caption{
(a) Experimental setup realizing passively phase-stable tunable lattices.
The lattice 2 beam (L2, blue) is out-of-plane polarized and forms a bow-tie lattice, realizing a square lattice potential (b).
The in-plane polarized lattice 1 beam (L1, orange) can be added with a well-defined superlattice phase $\Delta \varphi = 2 \pi \Delta x / (\lambda / 2)$ (see lattice potentials sketched in right inset).
For the in-phase case, $\Delta \varphi = 0$, this realizes an effective triangular lattice geometry (c).
(Left inset) The combined lattice is passively phase-stable due to the retro-reflection mirror serving as a common phase reference.
Temporal fluctuations of path lengths lead to translations along either common paths or along translationally invariant directions.
The arrows indicate the movement of the interference pattern generated by the respective lattice upon changes in the phase $\phi_{1, 2}$.
(d,e) Single-site resolved image of a Mott insulator in the square (d) and Lieb lattice (e).
(f) Lattices with more complex unit cells, e.g. as shown in (e), can be dynamically generated by projecting repulsive local potentials through the objective, blocking out distinct lattice sites.
(g-i) The MI phase hosts doublon-hole pairs (red shading), observable as correlated parities (blue: positive, grey: negative).
The brane parity serves as a proxy for a non-local order parameter and is given by the product of the on-site parities evaluated over an analysis area (black frame).
In the MI phase (g,h) its value is positive (for finite areas) and depends on the number of doublon-hole pairs  cut by the analysis boundary.
In the SF phase (i) parities are nearly uncorrelated, leading to a substantially smaller brane parity.
}
\label{fig:1}
\end{figure*}

\section{Programmable lattices}
In our approach to realizing tunable lattice geometries, we superpose a bow-tie lattice~\cite{sebby-strabley2006} (L2) with a mutually non-interfering retro-reflected 1d lattice (L1), see \figref[(a)]{fig:1}.
The two lattices are not only intrinsically phase-stable, but also relative to each other as they are both phase-referenced to a common retro-reflection mirror (see Appendix \ref{sec:lat-prop}).
The relative phase between the lattice potential minima, which we refer to as ``superlattice phase'' $\Delta \varphi$, can be adjusted by introducing a slight detuning between the lattice frequencies (taking into account the distance between atoms and retro-reflecting mirror).
By additionally varying the power ratio between the lattice beams $V_1 / V_2$, the ground band behavior can be tuned between square, triangular, honeycomb and 1d lattice, without the need for active phase locking.
On top of these base lattices, we employ a digital micromirror device (DMD) to project single-site-resolved beams through the microscope objective, see \figref[(f)]{fig:1}.
This procedure results in a programmable repulsive potential landscape, blocking atomic occupation on specific lattice sites, which allows for the realization of an even larger class of derived lattice potentials, see~\figref[(d,e)]{fig:1}.
At the same time, with light only applied to blocked out sites, this scheme minimizes cross-talk, reducing undesired local disorder.
The phase stability between these microscopic blocking beams and the base lattice is ensured by active feed-forward to correct for slow thermal drifts~\cite{weitenberg2011}.

\section{Lattice characterization}
In our experiment, we worked with about $200$ $^{87}\mathrm{Rb}$ atoms in the $\ket{F = 1, m_F = -1}$ ground state, trapped in lattices at a wavelength of $\lambda = \SI{1064}{\nano\meter}$ and with DMD block-out beams operating at $\SI{670}{\nano\meter}$.
For the data presented here, we optimized the superlattice phase for the triangular lattice condition $\Delta \varphi = 0$, and extracted a phase stability of $\sigma_{\Delta \varphi} = 0.01(1) \pi$ using L1 amplitude modulation spectroscopy (see Appendix \ref{sec:lat-prop}).
Starting with a 2d superfluid trapped in a single layer of a vertical 1d lattice, we adiabatically ramped up the horizontal lattices (L2, L1), such that the atoms formed a unity-filled Mott insulator with a typical filling of $0.97$.
After performing measurements in the desired lattice configuration, we ramped off L1 and performed single-site resolved fluorescence imaging in L2.
Due to pair-wise losses during fluorescence imaging, the resulting single-shot images reveal the local atom number parity~\cite{sherson2010}.

\begin{figure}
\centering
\includegraphics{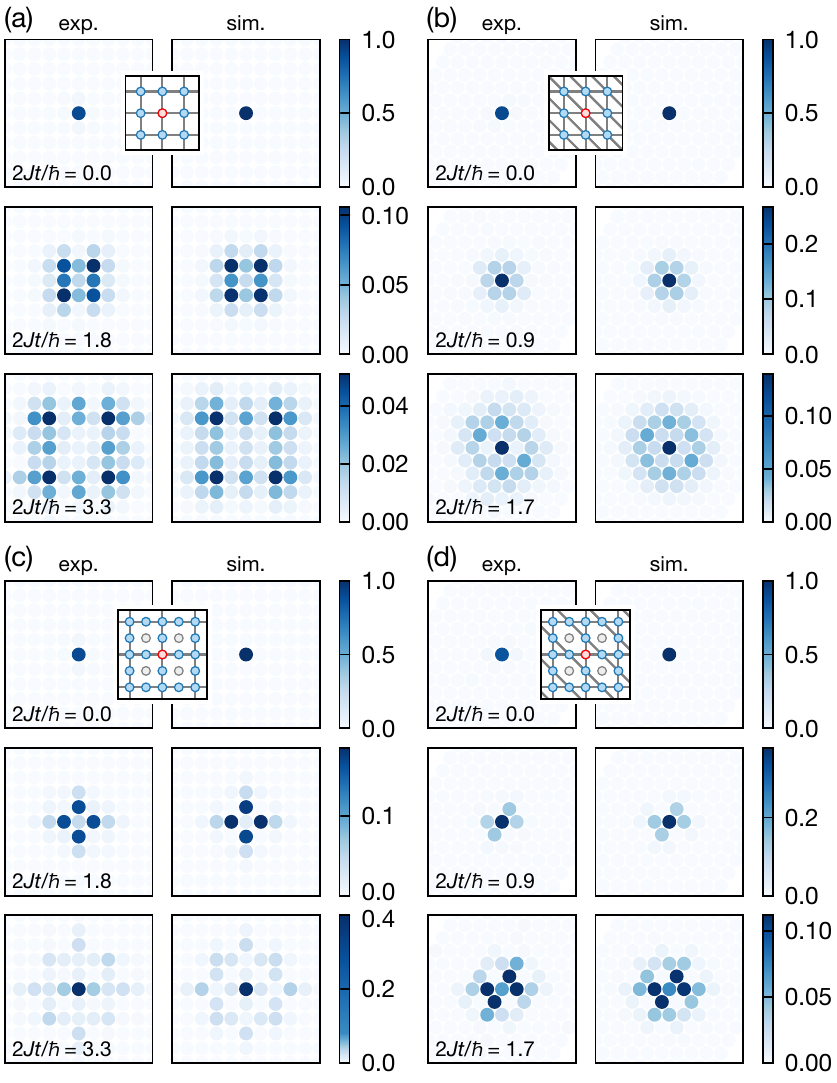}
\caption{
Atomic densities due to quantum walks in various lattice geometries.
After preparing a single localized atom in the center of the lattice (red site in insets), we measure the ballistic dynamics of the wavefunction at various times (top to bottom).
The square (a) and triangular (b) lattices are realized in the ground band of our superlattice.
The Lieb (c) and kagome (d) lattices are generated by locally projecting repulsive light on certain sites (grey sites in insets).
The interference fringes visible in the experimental data (left) agree well with simulations (right), indicating coherent evolution in a homogeneous and stable lattice.
Note that some color map ranges have been adjusted to facilitate displaying the large dynamic range.
}
\label{fig:2}
\end{figure}

To demonstrate the flexible control over the lattice geometry and benchmark the corresponding properties of the ground band, we perform single-particle quantum walks~\cite{fukuhara2013,preiss2015,young2022} in the respective 2d lattices.
To achieve this, we flip the hyperfine state of a single atom using our local microwave addressing technique based on a DMD~\cite{weitenberg2011,fukuhara2013}.
After pushing out all but the spin-flipped atom, we quench the lattices to a depth where the particle is allowed to tunnel.
As the wavefunction spreads coherently, we expect the evolving site-resolved probability distribution to display interference patterns characteristic for the specific lattice.

The density dynamics averaged over 250 experimental repetitions and its hopping symmetry axes is displayed in \figref{fig:2}, showing excellent agreement with simulations.
In the square lattice at $10.0(3) E_r^{(752)}$ depth, where $E_r^{(a / \si{\nano\meter})} = h^2 / 8 m a^2$ denotes the recoil energy of the respective lattice with spacing $a$, the two dimensions decouple and we observe the characteristic ballistically expanding wavefront with a fitted hopping energy of $J = h \times \SI{31(1)}{\hertz}$ along the horizontal and $J_v = 0.92(1) J$ along the vertical direction, see \figref[(a)]{fig:2}.
This agrees well with the hopping rates obtained from band structure calculations using the lattice depth independently calibrated by amplitude modulation spectroscopy.
The small observed anisotropy is well reproduced in our simulations when considering the difference in the lattice spacings as L2 intersects slightly non-orthogonally at an angle of $\SI{90.7(1)}{\degree}$.
For the triangular lattice, the depths are tuned to an isotropic coupling ratio, following the relation $V_1 / E_r^{(532)} \approx 4 + V_2 / E_r^{(752)}$.
The associated quantum walk measurements shown in \figref[(b)]{fig:2} were performed at $V_2 = 4.0(1) E_r^{(752)}$ with $J = h \times \SI{21(1)}{\hertz}$ and exhibit circularly symmetric expansion with a fitted residual diagonal anisotropy of $J_d = 1.05(2) J$.
In general, the tunability of the ratio $V_1 / V_2$ enables us to deliberately vary the diagonal anisotropy, interpolating between a square and a 1d lattice along the diagonal (see Appendix \ref{sec:single-particle}).

To characterize the emergent programmable lattices in presence of microscopic site-blocking potentials of $V_b = h \times \SI{300(90)}{\hertz}$, we measure quantum walks at the same base-lattice parameters as above.
For the block-out potential presented in \figref[(c,d)]{fig:2}, the expected lattice geometries are the Lieb or kagome lattices for the square or triangular base lattices, respectively.
We again find excellent agreement with simulations, and observe that the atom population remains on the non-blocked sites with $\SI{99(1)}{\percent}$ probability, while cross-talk-induced disorder is small.

\begin{figure}
\centering
\includegraphics{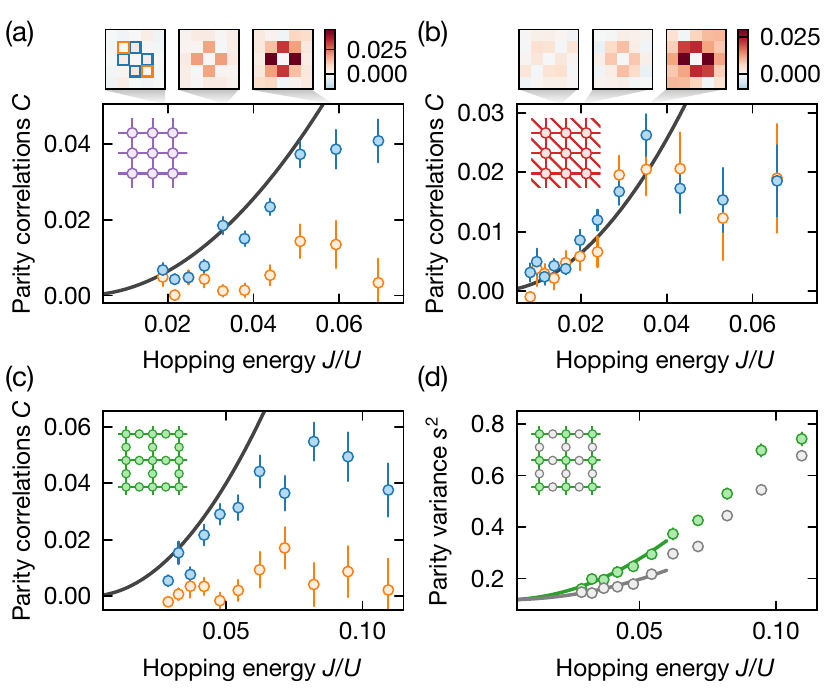}
\caption{
Doublon-hole fluctuations in the square (a) and the triangular lattice (b).
The straight-neighbor parity correlations (blue) grow with $(J / U)^2$ in both cases (line).
The correlations of the diagonal neighbor (orange) however only grow in the case of the triangular lattice.
The color plots (top) show the 2d parity correlations $C$ of the neighboring sites.
The colored edges in the left-most plot indicate the value shown in the main plot.
(c) The fluctuations in the Lieb lattice are averaged over both hub and rim sites and show a behaviour similar to the square lattice case in the perturbative regime $J \ll U$.
(Insets) in (a-c) depict the lattice geometry.
(d) Fluctuations are driven by coupling to neighboring sites, and thus the local coordination number.
The coordination number of the Lieb lattice depends on the site within the unit cell.
Accordingly, the on-site variance on the hub sites (green) grows twice as fast as the rim sites (grey), see inset.
Solid lines show perturbative calculations, with an offset that accounts for the finite filling of $0.97$.
Error bars denote the s.d. from a bootstrap analysis.
}
\label{fig:3}
\end{figure}

\section{Doublon-hole fluctuations}
After characterizing the single-particle tight-binding bands and the stability of the generated lattices through the quantum walks, we proceed to studying the interacting regime in the unity-filling Bose-Hubbard model realized on the various lattice geometries.
While the ground state in the atomic limit ($J / U\ll 1$) corresponds to a unity-filled product state, quantum fluctuations in the form of doublon-hole pairs emerge on top of the product state at finite tunnel couplings~\cite{endres2011,hartke2020}.
In a perturbative picture, regardless of the exact lattice geometry, every bond in an isotropic lattice is expected to give rise to equal nearest-neighbor $\avg{i, j}$ parity correlations of $C = \avg{\hat{s}_i \hat{s}_j} - \avg{\hat{s}_i} \avg{\hat{s}_j} \approx 16 J^2 / U^2$, where $\hat{s}_j = e^{i \pi (\hat{n}_j - 1)}$ denotes the local atom number parity.
In the experiment, we started with a 2d SF and then slowly ramped on the local block-out potential in $\SI{150}{\milli\second}$ to $V_b = h \times \SI{450(120)}{\hertz}$.
Subsequently, the horizontal lattices were adiabatically and isotropically increased to the depth corresponding to the desired $J / U$ parameters in $\SI{200}{\milli\second}$, followed by a fast $\SI{1}{\milli\second}$ ramp to $V_2 = 90 E_r^{(752)}$, which froze all quantum fluctuations.
The interaction energies in this measurement were in the range of $U = h \times 200 - \SI{300}{\hertz}$.
In \figref[(a-c)]{fig:3} we compare the correlations from 200 experimental runs evaluated over $9 \times 9$ sites along the straight and the diagonal neighbors for the square, triangular and Lieb lattice.
We clearly observe that diagonal correlations only arise in the case of the triangular lattice.
Furthermore, the growth in correlations agrees with the perturbative dependence within its range of validity for all lattice geometries along their respective bond directions, with deviations originating from hopping anisotropies, calibration uncertainties and finite temperatures.
When approaching the phase transition, the pairs rapidly deconfine, resulting in the observed reduction of neighboring correlations~\cite{endres2011}.

In the case of the tripartite Lieb lattice, there exist two types of sublattices with differing local coordination number $z$: the hub sites with $z = 4$ and the rim sites with $z = 2$.
This geometry gives rise to a flat central band, whose Bloch wavefunctions exclusively populate the rim sites~\cite{flannigan2021}, which suggests that the influence of the flat band might manifest as spatially distinct behavior on the two sublattice types.
In particular, in the SF phase the superfluid density is expected to be higher on the hub sites~\cite{rizzi2006,grygiel2022}, and may be viewed as a tendency to depopulate the flat band.
To capture the effects of this spatial inhomogeneity, we analyze the on-site variance $s^2 = \avg{\hat{s}_j^2} - \avg{\hat{s}_j}^2$ averaged over either sublattice type, see \figref[(d)]{fig:3}.
We can indeed observe that the variance differs between the two types of sites when approaching the phase transition, with the hub sites displaying higher fluctuations.
In the MI phase, the on-site fluctuations correspond to the formation of doublon-hole pairs with the site's $z$ neighbors and grow with $J / U$ as described by perturbation theory.
In the SF phase, we would similarly expect the sublattices to show distinct atom number fluctuations due to the inhomogeneous superfluid density.
However, as the parity is bounded, the parity variance is also bounded and at large $J / U$ the difference in the parity variance decreases again.

\begin{figure}
\centering
\includegraphics{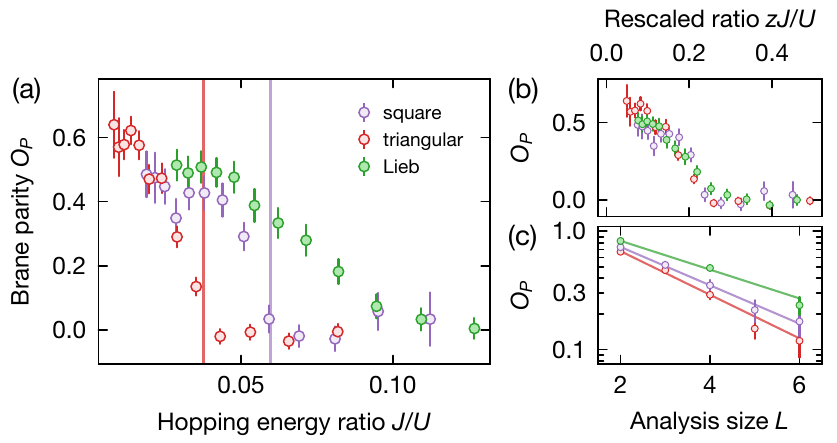}
\caption{
(a) Brane parity across the SF-MI phase transition for various lattice geometries analyzed over $4 \times 4$ sites.
The measurements in the triangular (red), square (purple) and Lieb (green) lattice all show a change from zero to finite values of the integer brane parity.
The critical $(J / U)_c$ agrees with the phase transition point obtained from quantum Monte-Carlo simulations (solid lines), indicating its suitability as non-local order parameter for the-Mott insulating phase.
(b) Rescaling the hopping energy with the respective (averaged) coordination number $z$ of the lattice, we find a collapse of the data, showing that the phase transition scales with $z$.
(c) Dependence of the integer brane parity with the analysis area containing $L \times L$ sites in the MI phase at $J / U = 0.029$ (triangular, red), $0.029$ (square, purple), and $0.033$ (Lieb, green).
Solid lines denote an exponential fit consistent with perimeter-law scaling.
The Lieb lattice only contains even data points due to its $2 \times 2$-site unit cell.
Error bars denote the s.d. from a bootstrap analysis.
}
\label{fig:4}
\end{figure}

\section{Brane parity}
The correlation properties of the doublon-hole pairs can furthermore be used to construct a non-local order parameter characterizing the Mott-insulating phase:
The value of the product of all parities within a region of interest, $\hat{O}_P = \prod_{i \in L \times L} \hat{s}_i$, is determined by the number of correlated doublon-hole pairs cut by its boundary.
Within the MI phase and for a finite brane, see \figref[(g,h)]{fig:1}, this amounts to a positive value of $O_P = \avg{\hat{O}_P}$, which decreases when approaching the phase transition due to the increased number of cut pairs and thus a higher probability for a negative $O_P$.
When reaching the SF phase, see \figref[(i)]{fig:1}, the proliferation of uncorrelated density fluctuations renders $O_P$ vanishing.
In the 1d Bose-Hubbard model, the string parity has been demonstrated to serve as a non-local order parameter, being finite in the MI phase and vanishing in the SF phase~\cite{berg2008,endres2011,rath2013}.
In 2d the integer brane parity would vanish also in the MI phase in the thermodynamic limit, and it was shown that using fractional instead of integer parities retains a non-vanishing order parameter~\cite{degliespostiboschi2016,fazzini2017}.
However, as $O_P$ decreases with a perimeter law~\cite{rath2013} and it remains finite for any \emph{finite} analysis area, the integer brane parity is still useful as a proxy for the true MI order parameter, and can capture the critical $(J / U)_c$ within experimental uncertainties.
In \figref[(a)]{fig:4}, we plot the integer brane parity evaluated over a $4 \times 4$ area as a function of $J / U$ for the triangular, square and Lieb lattice.
For data in the MI phase, varying the analysis area furthermore shows scaling consistent with a perimeter law, $\log O_P \sim -L$, see \figref[(c)]{fig:4}.
In all geometries, the location of the phase transition is clearly represented as a departure of the brane parity from zero.
The experimentally obtained critical values $(J / U)_c \approx 0.04$ and $0.06$ for the triangular and square lattice, respectively, agree well with quantum Monte-Carlo simulations~\cite{capogrosso-sansone2008,teichmann2010}.
We furthermore find a collapse of $O_P$ for the different lattices when rescaling the hopping energy with the coordination number of the lattice, see \figref[(b)]{fig:4}.
This scaling is consistent with predictions by mean field theory and measurements of the superfluid order parameter~\cite{thomas2017}, thus providing further validation for the use of the brane parity as a proxy for the non-local order parameter.
This scaling appears to also approximately hold true for the Lieb lattice at $(J / U)_c \approx 0.09$ when using the arithmetic mean of its local constituents as an effective coordination number.
The systematic deviations for the Lieb lattice towards higher parity values near the phase transition could hint at a stabilizing effect of the flat band on the MI phase, which is not captured by the applied simple rescaling with coordination number---a point that needs further investigation by theory and experiment.

\section{Conclusion}
Employing a quantum gas microscope, we have demonstrated a novel passively phase-stable approach to realizing 2d Hubbard systems in programmable lattice geometries.
In various of these lattices, our microscopic measurements have experimentally established the integer brane parity as a non-local observable suitable to characterize the 2d SF-MI phase transition.
Imaging within an out-of-phase superlattice would furthermore allow for resolving doublons~\cite{tarruell2012,greif2013} and could enable the detection of the fractional parity order for both bosonic and fermionic systems~\cite{degliespostiboschi2016,fazzini2017}.
Finally, our microscopic programmability of on-site potentials enables the exploration of further lattice-dependent many-body phenomena, ranging from the engineering of novel Hamiltonians on top of flat bands hosting exotic phases~\cite{diliberto2016,flannigan2021} to studying transport through interfaces between regions with differing lattice geometry.

\emph{Note added:}
During the completion of this manuscript, we became aware of related work studying fermionic many-body systems in actively phase-stabilized tunable lattice geometries~\cite{xu2022}.

\begin{acknowledgments}
We thank Jae-yoon Choi and Dan Stamper-Kurn for valuable discussions.
We acknowledge funding by the Max Planck Society (MPG), the European Union (PASQuanS grant no. 817482), and the Deutsche Forschungsgemeinschaft (DFG, German Research Foundation) under Germany's Excellence Strategy--EXC-2111--390814868.
K.S. and S.A. acknowledge funding from the International Max Planck Research School (IMPRS) for Quantum Science and Technology.
\end{acknowledgments}

\appendix

\section{Lattice properties}
\label{sec:lat-prop}

\subsection{Lattice phase stability}

In the tunable base lattices implemented in~\cite{tarruell2012,xu2022}, where two independent but mutually interfering retro-reflected laser beams are crossed, active phase stabilization is required due to the ``time phase'' difference between the two beams, $\alpha$, being an unrestricted degree of freedom.
For beams with wave number $k = 2 \pi / \lambda$ with a combined field given by $A \sim e^{i k y} + e^{-i k y} + e^{i \alpha} e^{i k x} + e^{i \alpha} e^{-i k x}$, the intensity becomes $|A|^2 \propto \cos 2 x + \cos 2 y + 4 \cos \alpha \cos x \cos y$ and thus realizes an interference structure that depends on $\alpha$.
As bow-tie lattices (as used in our setup) fold the same beam into the orthogonal axis, the time phase difference is inherently fixed to $\alpha = 0$.
In the following we furthermore show that this lattice is also structurally phase stable with respect to variations in the ``spatial phases'' when superposing with an additional 1d lattice.

\begin{figure}
\centering
\includegraphics{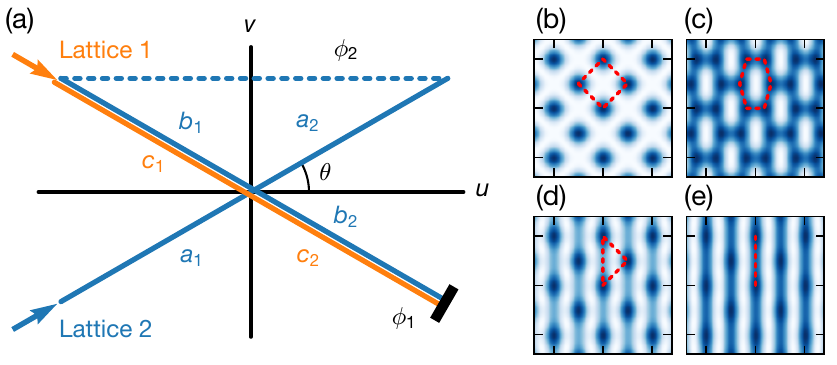}
\caption{
(a) Lattice beam layout denoting intersection half-angle $\theta$, field amplitudes of the incident beam passes $\{a_i, b_i, c_i\}$, and phase delays $\{\phi_i\}$ introduced by propagation.
The superlattice phase $\Delta \varphi$ is depicted in \figref[(a)]{fig:s:mod-spec}.
At an intersection angle of $2 \theta = \SI{90}{\degree}$, various lattice geometries can be realized, including:
in the absence of lattice 1 a square lattice (b),
in its presence a honeycomb lattice for $\Delta \varphi = \pi$ (c),
a triangular lattice for $\Delta \varphi = 0$ (d),
and a 1d lattice in the limit of deep lattice 1 (e).
}
\label{fig:s:beam-layout}
\end{figure}

The layout of our lattice beams is shown in \figref{fig:s:beam-layout} with the two axes $\mathbf{k}_{x, y} = k (\cos \theta, \mp \sin \theta)$.
The square lattice generated by lattice 2 and the 1d lattice generated by lattice 1 have respective fields of
$
A_2 = a_1 e^{i (\mathbf{k}_y \cdot \mathbf{r})}
+ b_1 e^{i (\mathbf{k}_x \cdot \mathbf{r} + \phi_2)}
+ b_2 e^{i (-\mathbf{k}_x \cdot \mathbf{r} + \phi_2 + 2 \phi_1)}
+ a_2 e^{i (-\mathbf{k}_y \cdot \mathbf{r} + 2 \phi_2 + 2 \phi_1)}
$
and
$
A_1 = c_1 e^{i (\mathbf{k}_x \cdot \mathbf{r})}
+ c_2 e^{i (-\mathbf{k}_x \cdot \mathbf{r} + 2 \phi_1 + \Delta \varphi)},
$
where $\Delta \varphi$ indicates the superlattice phase.
This yields an overall light intensity of
\begin{align}
I   &= |A_1|^2 + |A_2|^2 \nonumber \\
&= (a_1^2 + a_2^2 + b_1^2 + b_2^2 + c_1^2 + c_2^2) \nonumber \\
&+ 2 c_1 c_2 \cos[2 k (u - u_0) \cos \theta - 2 k (v - v_0) \sin \theta - \Delta \varphi] \nonumber \\
&+ 2 a_1 a_2 \cos[2 k (u - u_0) \cos \theta + 2 k (v - v_0) \sin \theta] \nonumber \\
&+ 2 b_1 b_2 \cos[2 k (u - u_0) \cos \theta - 2 k (v - v_0) \sin \theta] \nonumber \\
&+ 2 (a_1 b_1 + a_2 b_2) \cos[2 k (v - v_0) \sin \theta] \nonumber \\
&+ 2 (a_1 b_2 + a_2 b_1) \cos[2 k (u - u_0) \cos \theta]
\label{eq:s:lat-intensity}
\end{align}
where we have defined $2 k u_0 \cos \theta = \phi_2 + 2 \phi_1$ and $2 k v_0 \sin \theta = \phi_2$.
Thus, the lattice potential only depends on a translated position $(u - u_0, v - v_0)$, confirming that the lattice is structurally passively phase stable.

\subsection{Bose-Hubbard parameters}

The Bose-Hubbard model is given by
\begin{equation}
\hat{H} = -\sum_{\avg{i, j}} J_{ij} \hat{c}_i^\dagger \hat{c}_j
+ \frac{U}{2} \sum_i \hat{n}_i (\hat{n}_i - 1)
+ \sum_i V_i \hat{n}_i.
\end{equation}
The on-site potential $V_i$ was experimentally calibrated, as described in Appendix~\ref{sec:many-body}.
The hopping energy $J_{ij}$ and interaction energy $U$ were theoretically calculated and verified for certain values by fitting the quantum walk measurements and modulation spectroscopy, respectively.
Since the lattice potential $V (u, v) \propto -I (u, v)$ is not separable, we perform a full 2d band structure calculation following~\cite{bissbort2012}.
On the one hand, this calculation yields the band gaps used for the lattice depth calibration (see Appendix~\ref{sec:single-particle}).
On the other hand, we obtain the ground state Wannier wavefunctions $w_j (u, v)$ on lattice site $j$, which we use to determine the hopping energy between sites $i$ and $j$ by evaluating
\begin{equation*}
J_{ij} = \int du \hspace{2pt} dv \hspace{2pt} w_i^* (u, v) \left(-\frac{\hbar^2}{2 m} \nabla^2 + V (u, v) \right)w_j (u, v)
\end{equation*}
and the Hubbard interaction energy
\begin{equation*}
U = \frac{4 \pi \hbar^2 a_s}{m} \int du \hspace{2pt} dv \hspace{2pt} dz \left|w (u, v) w^z (z) \right|^4
\end{equation*}
where $a_s$ is the $s$-wave scattering length.
The Wannier function for the vertical direction $w^z (z)$ is independently obtained from a 1d band structure calculation due to the separability of the lattice potential along this direction.

For the lattice geometries with site block-out, we consider the influence of the block-out potentials within the tight-binding model since the band gaps of $\gtrsim h \times \SI{3}{\kilo\hertz}$ are much larger than the block-out potentials of $\lesssim h \times \SI{450}{\hertz}$.

\section{Single-particle measurements}
\label{sec:single-particle}

\subsection{Modulation spectroscopy}

We calibrate the individual lattice depths by performing amplitude modulation spectroscopy and find two $d$-band resonances from which we determine the lattice depth with $\sim \SI{2}{\percent}$ uncertainty.

\begin{figure}
\centering
\includegraphics{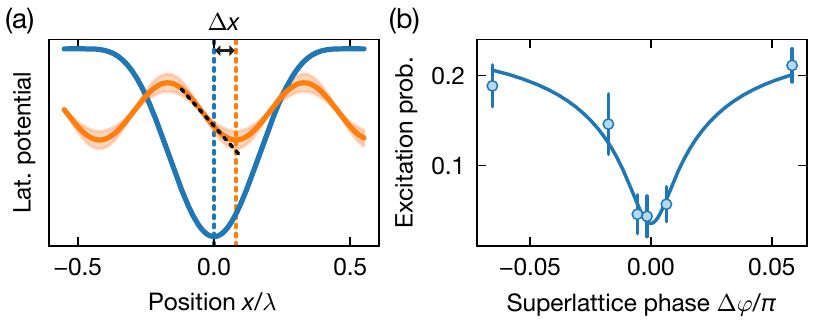}
\caption{
(a) The superlattice phase, $\Delta \varphi = 2 \pi \Delta x / (\lambda / 2)$, can be precisely calibrated by amplitude-modulating lattice 1 (orange solid) near the band gap frequency of a much deeper lattice 2 (blue solid).
Single-band excitations require dipolar modulation (black dashed), which is minimal when the lattices are in phase ($\Delta \varphi = 0$).
The vertical dashed lines represent the potential minima (and thus phase) of the respective lattices.
(b) Single-band amplitude modulation spectroscopy probing the resonant band excitation probability at the respective superlattice phase.
The solid line shows a fit from which we extract a superlattice phase stability of $\sigma_{\Delta \varphi} = 0.01(1) \pi$, which was confirmed in a long-time measurement.
}
\label{fig:s:mod-spec}
\end{figure}

For the superlattice phase measurements shown in \figref{fig:s:mod-spec}, we amplitude-modulate lattice 1 within a deep lattice 2 potential near the upper $p$-band resonance.
Due to the weak single-particle drive in an isolated system, we analyze the response assuming a two-level model with coupling $\Omega$ and modulation-frequency detuning $\Delta$.
This model yields a mean excited-state population of $P_e (\Omega, \Delta) = 2 / (4 + \delta^2 + \sqrt{\delta^2 (4 + \delta^2)})$, with $\delta = \Delta / \Omega$.
Close to the superlattice in-phase condition, $\Delta \varphi = 0$, the coupling is proportional to the superlattice phase (here: $\Omega / \Delta \varphi \approx \SI{660}{\hertz}~h / \pi$), which we calculate from the band structure results.
Considering the long and weak drive, we assume that Gaussian fluctuations in the superlattice phase $(\propto \sigma_\Omega)$ and in the lattice depth $(\propto \sigma_\Delta)$ dominate the shape of the resonance.
Thus, the excitation probability on resonance is given by the two-fold convolution over the fluctuations, yielding
\begin{equation}
\overline{P}_e (\Omega)
\sim 1 - \int f_{\mathcal{N} (\omega, \sigma^2)} (x)
e^{x^2} \mathrm{erfc} |x| dx,
\label{eq:s:mod-spec}
\end{equation}
where $\mathrm{erfc}$ denotes the complimentary error function, and $f_{\mathcal{N} (\omega, \sigma^2)}$ the probability distribution of a normal distribution with center $\omega^2 = 2 \Omega^2 / \sigma_\Omega^2$ and variance $\sigma^2 = \sigma_\Omega^2 / 2 \sigma_\Delta^2$.

In the experiment, we tuned the superlattice phase by varying the frequency difference $\Delta f$ between the lattices by an acousto-optic modulator, which yields a tuning slope of $\Delta \varphi / \Delta f \approx \pi / \SI{250}{\mega\hertz}$ for the distance between atoms and retro-reflecting mirror of $\sim \SI{300}{\milli\meter}$.
At a lattice 2 depth of $V_2 = 185(5) E_r^{(752)}$, where all dynamics in the lattice is frozen and where we can separate the lattice in its local potential wells, we modulated lattice 1 at $V_1 = 5.0(2) E_r^{(532)}$ with a modulation depth of $0.25$.
As a spectroscopic signature, we measured the fraction of atoms remaining in the ground band after modulation by adiabatically lowering the lattice depth to $V_2 \approx 18 E_r^{(752)}$, leading to the loss of atoms populating higher bands, see \figref[(b)]{fig:s:mod-spec}.
By fitting the functional shape of \eqnref{eq:s:mod-spec} to our experimental data and converting from coupling strength to the superlattice phase, this model allows us to extract a standard deviation of the superlattice phase of $\sigma_{\Delta \varphi} = 0.01(1) \pi$.
Repeating this measurement weeks later gave a similar excitation probability, demonstrating the long-term stability of this lattice scheme.

\subsection{Quantum walks}

The quantum walk measurements shown in \figref{fig:2} were performed by preparing a single atom and lowering the lattice 2 depth from $25 E_r^{(752)}$ (for square geometries) and $15 E_r^{(752)}$ (for triangular geometries) to the depth used for the dynamics measurements in $\SI{2.5}{\milli\second}$ to avoid band excitations.
After the time evolution, we froze the dynamics by ramping up lattice 2 to $25 E_r^{(752)}$ in $\SI{0.8}{\milli\second}$.
We postselected the data for a single detected atom and fitted the resulting time-dependent densities to numerical simulations of the respective tight-binding lattice.
For the base lattice geometries, we fitted the hopping energy along each bond direction, as well as a time offset $t_0$ to account for the finite ramp times, yielding $2 J t_0 = 0.60(1)$ and $0.32(5)$ for the square and the triangular geometries, respectively.
The parameters of the base lattice fits were used for the Lieb and kagome lattice simulations and agreed with a direct fit to the data.

\begin{figure}
\centering
\includegraphics{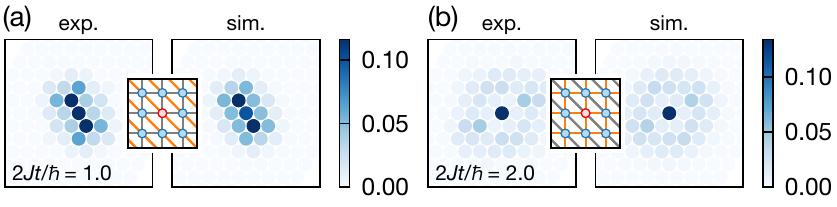}
\caption{
Densities due to quantum walks in anisotropic triangular lattices with fitted hopping energy ratios between diagonal and horizontal neighbors of $J_d / J = 1.69(3)$ (a) and $0.79(2)$ (b).
(Insets) show the site connectivity with stronger couplings highlighted in orange.
}
\label{fig:s:qw-aniso}
\end{figure}

By varying the depth ratio $V_1 / V_2$ between the lattices, we can furthermore tune the hopping energy ratio between the straight bonds and the diagonal bonds, i.e., the geometry between a square lattice for $V_1 \ll V_2$
and a 1d lattice for $V_1 \gg V_2$.
In measurements similar to the ones shown in \figref{fig:2} at lattice depths of $V_2 = 3.9 E_r^{(752)}$ and $V_1 \in \{5.9, 9.4\} E_r^{(532)}$, we performed quantum walks subject to intermediate anisotropic hopping ratios, see \figref{fig:s:qw-aniso}.
Fitting to numerical simulations yields $J_d / J = 1.69(3)$ and $0.79(2)$, which similarly shows good agreement between simulations and calibrations, demonstrating control of the anisotropy.

\section{Many-body measurements}
\label{sec:many-body}

\subsection{On-site potential calibration}

Our vertical lattice creates a spatially inhomogeneous in-plane confinement potential.
To estimate its potential depth, we increase the atom number loaded into the system until a doubly-filled Mott insulator forms in the center.
The outline of the atomic cloud then gives us the equipotential line at a potential depth of the Hubbard interaction energy $U$.
Due to fluctuations in the atom number, the major source of uncertainty for this calibration method stems from determining the outline of the cloud.

Using this information, we calibrated the projected DMD potential for blocking out the lattice sites.
We adiabatically ramped the square lattice into the atomic limit with the projected potential switched on.
While keeping the atom number such that the outline of the atomic cloud remained near the $U$-equipotential line, we varied the projected light power.
When reaching a projected potential of $U$, we expect the population on the central blocked-out sites to vanish.
We therefore calibrated the DMD potential by mapping the light power where the average filling of the central blocked-out sites reached $\lesssim 0.03$ to a potential shift of $\sim U$.

\subsection{Lieb sublattice inhomogeneity}

\begin{figure}
\centering
\includegraphics{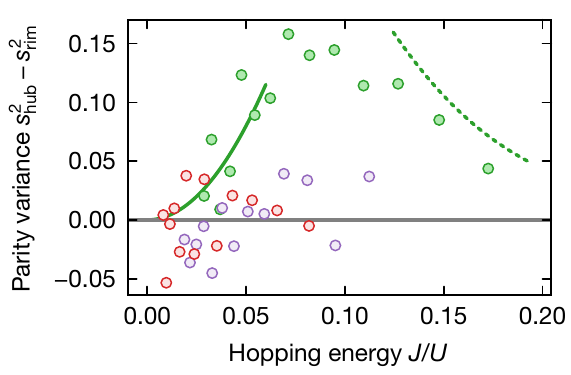}
\caption{
Parity variance on the hub sites $s_\mathrm{hub}^2$ subtracted by the variance on the rim sites $s_\mathrm{rim}^2$ for the triangular (red), square (purple) and Lieb (green) lattices.
The local variance differs significantly only for the Lieb lattice.
The solid line indicates perturbative on-site fluctuations from doublon-hole pairs in the MI phase.
The dashed line indicates inhomogeneous mean-field calculations at $\mu / U = 0.5$ in the SF phase.
}
\label{fig:s:loc-var-dev}
\end{figure}

In order to validate the observation that the parity variance differs between the hub and rim sublattice sites of the Lieb lattice as shown in \figref[(d)]{fig:3}, we plot the variance difference in \figref{fig:s:loc-var-dev}.
Comparing to the same analysis performed in the square and triangular lattice, we can see that only the Lieb lattice shows a significant deviation from zero.

We would expect the sublattice-dependent occupation fluctuations to grow further into the SF phase, however, the data show a peak already around the phase transition point.
We attribute this observation to the fact that, in contrast to the atom number, the parity is bounded, which limits the observable fluctuations.
This behavior is also qualitatively reproduced by inhomogeneous mean field calculations~\cite{rizzi2006}, which similarly show a reduction in the parity variance difference with increasing $J / U$.

\subsection{Finite-size scaling of the integer brane parity}

\begin{figure}
\centering
\includegraphics{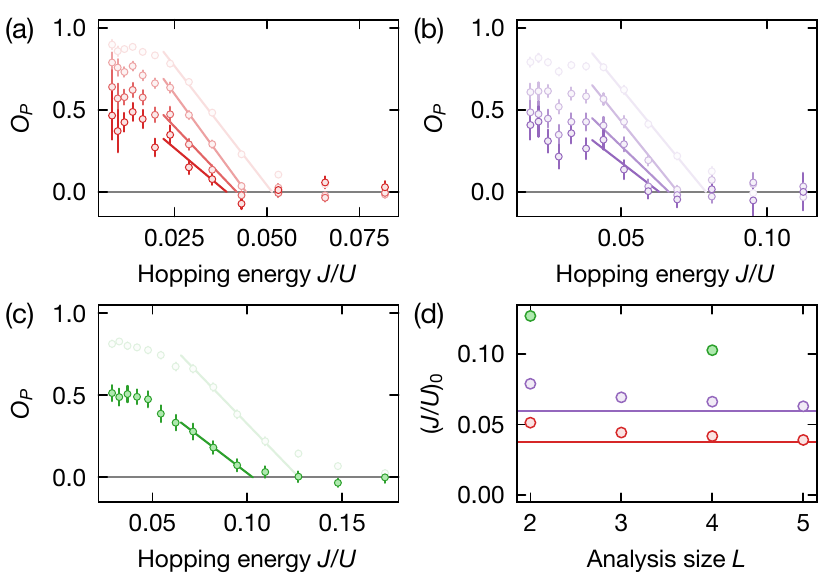}
\caption{
(a-c) Brane parity in the triangular (red), square (purple) and Lieb (green) lattice for increasing analysis sizes (light to dark color), ranging from $L = 2 - 5$ for the triangular and square lattice, and $L = 2, 4$ for the Lieb lattice.
The lines indicate linear fits to the sloped $J / U$-parameter regime.
We extract a simplified estimate for the critical point, $(J / U)_0$, as the point where the fit vanishes.
(d) With increasing analysis size $L$, the extracted phase transition point converges.
The horizontal lines indicate the critical value predicted by quantum Monte Carlo simulations.
}
\label{fig:s:bp-steepness}
\end{figure}

To maximize the signal-to-noise ratio of the integer brane parity extracted from experimental data, we first crop the images to a $7 \times 7$-site area in the center of the atomic cloud.
We then evaluate the brane parity for all possible $L \times L$-site analysis areas within the original $7 \times 7$ sites and average over the results.
Note that in the case of the Lieb lattice, we flip the sign of $O_P$ for analysis areas with an odd number of total lattice sites.
In this section, we will discuss how the choice of $L$ influences the value of $O_P$ as well as the extracted phase transition point.

In the MI phase, we show that the integer brane parity $O_P$ is subject to a perimeter-law scaling, $\log O_P \sim -L$, see \figref[(c)]{fig:4}.
Evaluated at different parameter regimes of $J / U$, with increasing $L$ we additionally observe a slight trend towards lower values than expected for a perimeter law.
This behaviour can be partially attributed to finite-temperature effects, which lead to the formation of uncorrelated individual holes.
Uncorrelated holes would lead to an area-law scaling, $\log O_P \sim -L^2$, and thus a downward trend that becomes more dominant with increasingly large analysis sizes (due to the perimeter-area scaling) and with decreasing $J / U$ (due to the reduced probability of finding correlated pairs).
Another reason involves the inhomogeneous confining potential from the lattice beams, leading to a coexistence of different phases in the system depending on the local chemical potential~\cite{sherson2010}.
As a consequence, we expect a bias towards a superfluid when including regions of smaller local chemical potential towards the edges of the atomic cloud.
As analyzing with larger $L$ has a higher sampling frequency at the edges than with smaller $L$, the inhomogeneity effects are stronger for larger $L$.

In the SF phase one would in contrast expect $\log O_P \sim -L \log L$ scaling~\cite{rath2013}.
We do not directly observe such scaling since the absolute $O_P$ values are much smaller and lie within experimental noise already at $L \sim 4$.
However, due to the difference in scaling compared with the MI phase, we expect the integer brane parity to serve as a more accurate proxy for the order parameter when measured on larger analysis areas $L \times L$.
In \figref{fig:s:bp-steepness} we show the $J/U$-dependence of $O_P$ for different $L$ and extract a simplified estimate for the phase transition point $(J / U)_0$:
We linearly fit the sloped part of the data (disregarding nonlinear behavior predicted in the immediate vicinity of the phase transition~\cite{endres2011,rath2013}) and assign the value at which the fit vanishes as $(J / U)_0$, for which we indeed observe convergent behavior for increasing $L$.

\end{document}